\documentclass[twoside,letterpaper]{article}
\usepackage[utf8]{inputenc}
\usepackage[english]{babel}
\usepackage[sorting=none,style=nature,doi=false,isbn=false,url=false,backend=biber]{biblatex}
\usepackage{csquotes}
\usepackage{amsmath}
\usepackage{amssymb}
\usepackage{graphicx}
\usepackage{color}
\usepackage{float}
\usepackage{ragged2e}
\usepackage{breakcites}
\DeclareGraphicsExtensions{.pdf,.png,.jpg}
\usepackage{wrapfig}
\usepackage{multicol}
\usepackage{fancyhdr}
\usepackage{helvet}
\usepackage{setspace}
\usepackage[lmargin=0.7in,rmargin=0.7in,tmargin=1in,bmargin=1in]{geometry}
\usepackage[labelfont={bf},labelsep=period,justification=justified,font=small,skip=0pt]{caption}
\usepackage{hyperref}
\hypersetup{colorlinks=true,citecolor=blue,linkcolor=black,urlcolor=blue}
\bibliography{main}
\usepackage[dvipsnames]{xcolor}
\usepackage{titlesec}
\titleformat*{\section}{\large\bfseries\sffamily\color{BrickRed}\fontencoding{T1}\fontfamily{phv}\selectfont}
\titleformat*{\subsection}{\fontencoding{T1}\fontfamily{phv}\bfseries}
\titleformat*{\subsubsection}{\fontencoding{T1}\fontfamily{phv}\itshape}
\usepackage{textgreek} 
\begin{document}
\pagestyle{fancy}
\fancyhf{}
\renewcommand{\headrulewidth}{0pt}
\renewcommand{\footrulewidth}{0pt}
\fancyfoot[RO]{Hsu \textit{et al.}, PREPRINT $|$ \textbf{\thepage}}
\fancyfoot[LE]{\textbf{\thepage} $|$ Hsu \textit{et al.}, PREPRINT}
\singlespacing

\vspace*{2cm}
\begin{LARGE}
\fontencoding{T1}\fontfamily{phv}
\noindent\textbf{Numerical parameter space compression and its application to microtubule dynamic instability}
\end{LARGE}

\vspace{15pt}
\noindent {\fontencoding{T1}\fontfamily{phv}\selectfont Chieh-Ting (Jimmy) Hsu\textsuperscript{a}, Gary J. Brouhard\textsuperscript{b,a*}, Paul François\textsuperscript{a,b*}}

\vspace{5pt}
\begin{scriptsize}
\fontencoding{T1}\fontfamily{phv}\selectfont
\noindent\textsuperscript{a}Department of Physics; \textsuperscript{b}Department of Biology, McGill University, Montréal, QC, Canada
\end{scriptsize}

\begin{scriptsize}
\fontencoding{T1}\fontfamily{phv}\selectfont
\noindent\textsuperscript{*}To whom correspondence should be addressed. E-mail: gary.brouhard@mcgill.ca, paul.francois2@mcgill.ca
\end{scriptsize}

\section*{ABSTRACT}
\vspace{-10pt}
\noindent{\fontencoding{T1}\fontfamily{phv}\selectfont Physical models of biological systems can become difficult to interpret when they have a large number of parameters. But the models themselves actually depend on (\emph{i.e.} are sensitive to) only a subset of those parameters. Rigorously identifying this subset of ``stiff'' parameters has been made possible by the development of parameter space compression (PSC). However, PSC has only been applied to analytically-solvable physical models. We have generalized this powerful method by developing a numerical approach to PSC that can be applied to any computational model. We validated our method against analytically-solvable models of random walk with drift and protein production and degradation. We then applied our method to an active area of biophysics research, namely to a simple computational model of microtubule dynamic instability. Such models have become increasingly complex, perhaps unnecessarily. By adding two new parameters that account for prominent structural features of microtubules, we identify one that can be ``compressed away'' (the ``seam'' in the microtubule) and another that is essential to model performance (the ``tapering'' of microtubule ends). Furthermore, we show that the microtubule model has an underlying, low-dimensional structure that explains the vast majority of our experimental data. We argue that numerical PSC can identify the low-dimensional structure of any computational model in biophysics. The low-dimensional structure of a model is easier to interpret and identifies the mechanisms and experiments that best characterize the system.}

\vspace{10pt}
\begin{scriptsize}
\fontencoding{T1}\fontfamily{phv}\selectfont
\noindent Numerical parameter space compression $|$ Fisher information matrix $|$ Stochastic processes $|$ Monte Carlo simulations $|$ Biological modeling $|$ Microtubule dynamics
\end{scriptsize}

\begin{multicols}{2} 
\vspace{-10pt}
\section*{INTRODUCTION}
\vspace{-10pt}
A central goal of biophysics is to develop mathematical and computational models that describe biological systems. These models can operate at different temporal and spatial scales. In the case of the microtubule cytoskeleton, models range from molecular dynamics simulations of \textalpha\textbeta-tubulin heterodimers~\cite{Mitra_Sept_2008}, to Monte Carlo simulations of microtubule dynamic instability~\cite{VanBuren_Odde_2002,VanBuren_Odde_2005,Castle_Odde_2017}, to analytical theories that treat the mitotic spindle as a nematic liquid crystal~\cite{Lydon_2006}. These models vary in their degree of complexity, e.g., in the number of parameters they use.

A central problem in biophysical modeling is defining the ``right'' number of parameters to explain and predict experimental data, or observables. We prefer simple models; in the well-known quip from Von Neumann, four parameters are sufficient to fit an elephant, and five can make its trunk wiggle~\cite{Dyson_Freeman_2004}, as was indeed later demonstrated~\cite{Mayer_Howard_2010}. More parameters can sometimes improve a model's performance, but too many can be a problem. Unnecessary parameters obfuscate those that determine the model's output, namely its reproduction of observables, and render the model less interpretable---all without any gain in predictive power~\cite{Gunawardena_Jeremy_2014}. A recent computational model of microtubule dynamic instability has 22 parameters and reproduces an impressive range of experimental data~\cite{Zakharov_Grishchuk_2015}. But complex models can be black boxes; we need a rigorous way to define which parameters determine, for example, the distribution of microtubule lifetimes.

The behavior of a model can be described within a so-called ``parameter space'', which has as many dimensions as there are parameters. Moving within this parameter space (by changing the values of parameters) should change a model's output. But usually a given observable significantly changes along only a few directions in parameter space~\cite{Gutenkunst_Sethna_2007}. In other words, most directions in parameter space are irrelevant. In order to make sense of complex models, an important scientific problem is to reliably extract relevant directions in parameter space, defining the true, lower-order ``dimensionality'' of the model. There are several methods to solve this problem. In the 1980's, classical Principal Component Analysis was proposed as a method to reduce ODE-based models of biochemical systems~\cite{Vajda_1985}. More recently, the Manifold Boundary Approximation Method has been developed to fit data while minimizing dimensionality~\cite{Transtrum_Qiu_2014}; similarly, Fitness Based Asymptotic Parameter Reduction can extract the ``core working module'' of a model~\cite{Proulx_Giraldeau_Rademaker_Francois_2017}. Other machine learning approaches can develop realistic models with a minimal number of parameters, e.g., using Bayesian Information Criterion~\cite{Daniels_Nemenman_2015}. These methods are focused on ODE-based models, so there is an acute need for universal methods that are applicable to stochastic computational models as well.

Our method to reduce the complexity of stochastic computational models is parameter space compression (PSC)~\cite{Machta_Sethna_2013}. The PSC method uses the Fisher Information Matrix (FIM) in order to determine the relative significance of a model's parameters. More specifically, the PSC method tracks the eigenvalues of the FIM over time to identify combinations of parameters that are ``stiff'' (\emph{viz.}, those with strong effects on model outputs) and ``sloppy'' (those with very weak effects) (Fig.~\ref{fig:Method}A) ~\cite{Transtrum_Sethna_2011}. The sloppy parameters or parameter combinations are ``compressed away'' to reveal the simpler dimensionality that underlies a model's performance~\cite{Transtrum_Sethna_2015}. That most parameters are sloppy, and thus irrelevant, is why coarse-grained models in physics provide such satisfying descriptions of the natural world~\cite{Waterfall_Sethna_2006}.

PSC is incredibly powerful at identifying the low-dimensional structure of a model~\cite{Machta_Sethna_2013}, but it has currently been applied to a limited number of analytically-solvable physics models. Those derivations are not easy to generalize to other contexts. In order to apply PSC to any computational model, we developed a numerical PSC method, significantly expanding the applicability of PSC. To validate our method, we recovered the analytical results of a simple one-dimensional random walk model~\cite{Machta_Sethna_2013} and its perturbations. We further tested our method on an analytically-solvable model of protein production and degradation. Finally, we applied numerical PSC to a well-known Monte Carlo model of microtubule dynamic instability. For the microtubule case, we identify a low-dimensional description of the system that accounts for all of our observables. By adding new parameters to our base model, we demonstrate that a feature of microtubules of current interest~\cite{Zhang_Nogales_2018}, namely the ``seam'' in the lattice, is ``compressed away'', while a feature neglected in some models, namely the fine structure of microtubule ends, is critical to model performance. In all three test cases, we show that the eigenvalues of the FIM provide critical insights into the behavior of a model and the importance of its parameters. Thus, our numerical PSC method opens the door to an analysis of computational models in biophysics that reveals the minimal yet predictive descriptions of living systems.
\vspace{-10pt}
\section*{NUMERICAL PARAMETER SPACE COMPRESSION}
\vspace{-10pt}
\subsection*{Mathematical formulation}
\begin{figure*}[!htb]
\centering
\includegraphics[width=1\textwidth]{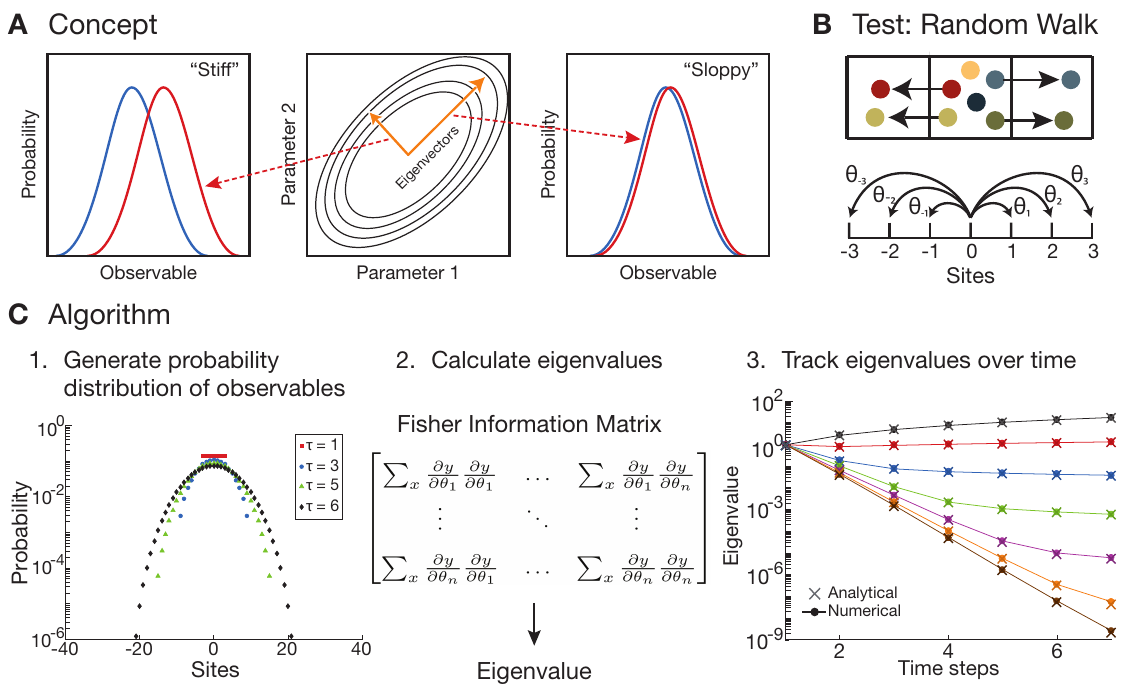}
\caption{\fontencoding{T1}\fontfamily{phv}\selectfont\textbf{Numerical PSC.} (A) Schematic of a (``stiff'') parameter versus a (``sloppy'') parameter in parameter space. The ``stiff'' parameter changes the observable more significantly than a ``sloppy'' one. Note: the red and blue curves are probability distributions when shifting different parameters by the same amount. (B) Schematic of one-dimensional random walk with parameters $\theta_{i}$, the probability of jumping to neighboring sites. (C) The three steps of our numerical PSC method: 1. Generate the probability distributions of the observables needed (particle density at different time steps), 2. calculate the finite derivatives for the Fisher Information Matrix and its corresponding eigenvalues, and 3. repeat at each time step of the simulation and track the eigenvalues over time. Note: our numerical PSC is able to reproduce the analytic result from~\cite{Machta_Sethna_2013}.}
\label{fig:Method}
\end{figure*}

\vspace{-5pt}
Our numerical approach is derived from the theoretical work of Machta and colleagues and relies on the computation of the Fisher Information Matrix (FIM)~\cite{Machta_Sethna_2013,Transtrum_Sethna_2011,Transtrum_Sethna_2015}. We study the distribution of an observable $x$ and its sensitivity to a vector of parameters $\vec{\theta}=\{\theta_\mu\}$. The central idea is that changes in an important parameter will result in bigger changes in the probability distributions $y(\vec{\theta},x)$ compared to changes in a less important one 
(Fig.~\ref{fig:Method}A). The important parameters will dominate the eigenvalues and eigenvectors of the FIM. Dominating eigenvalues define directions in parameter space where observables vary significantly. We call these directions ``effective parameters.'' In general, the effective parameters of a model are not the original parameters but rather combinations of them (see below). Thus, the goal of parameter space compression (PSC) is to identify these dominating eigenvalues and eigenvectors, which will define the most important directions in parameter space and the effective parameters defining the distribution $y(\vec{\theta},x)$ of observable $x$~\cite{Transtrum_Sethna_2011,Transtrum_Sethna_2015}. In particular, for a dynamical system, we expect that  a hierarchy of eigenvalues will appear for the FIM of $y(\vec{\theta},x,t)$ as time $t$ progresses, such that only a few effective parameters define the observed dynamics at any given time. These few effective parameters are sufficient to completely describe the system.

The FIM can be difficult to compute, but a simplification arises if we assume that the deviations of the noisy function $y$ are Gaussian distributed to the true model. As shown in~\cite{Machta_Sethna_2013}, the FIM at any given time $t$ can then be rewritten as a simple \emph{deterministic} ``metric'':
\begin{equation}
g_{\mu,\nu}(t) = \sum_{x} \frac{\partial y(\theta,x,t)}{\partial\theta_{\mu}} \frac{\partial y(\theta,x,t)}{\partial \theta_{\nu}}
\label{eq:FIM}
\end{equation}
where $y$ can be evaluated as a function of time $t$. Notice that $g_{\mu,\nu}(t)$ then becomes a simple function of the Jacobian with respect to its parameters. A detailed derivation is provided in Appendix S$1$.
We independently consider several observables $x$, and for each $x$ compute the FIM of distribution $y(\vec{\theta},x, t)$ and its eigenvalues as a function of time $t$.

\vspace{-10pt}
\subsection*{Scaling and Algorithm}
\vspace{-5pt}
A challenge in analyzing computational models, especially in biology, is that the parameters have different units and scales. Some parameters are energies (e.g., the $\Delta G^{o}$ of bond formation) and some are kinetic rate constants (e.g., the rate constant of a GTP hydrolysis reaction). Since rate constants are exponentially distributed to thermal energy $k_{B}T$, we choose to rescale the parameters to express all of them in terms of energies when calculating the FIM. Energies are more fundamental quantities and their variations are easier to interpret physically. We thus define newly rescaled parameters, $\tilde \theta_\mu$, so that $\tilde \theta\mu= \theta_\mu$ for parameters that are energies already, and  $\tilde \theta\mu=\log \theta_\mu$ for  rate constants (with an implicit conversion factor to remove units), as done previously~\cite{Gutenkunst_Sethna_2007}. Therefore, equation~\ref{eq:FIM} becomes:
\begin{equation}
g_{\mu,\nu} = \sum_{x} \frac{\partial y}{\partial \tilde \theta_{\mu}} \frac{\partial y}{\partial \tilde \theta_{\nu}}= \sum_{x} \frac{\partial y_{i}}{\partial \theta_{\mu}} \frac{\partial y_{i}}{\partial \theta_{\nu}}\theta_{\mu}^{\alpha_\mu}\theta_{\nu}^{\alpha_\nu}
\label{eq:FIM_log}
\end{equation}
where  $\alpha_\mu=0$ if $\theta_\mu$ is an energy and $\alpha_\mu=1$ if $\theta_\mu$ is a rate constant.

To calculate the FIM numerically using equation~\ref{eq:FIM_log}, we developed a three-step algorithm shown in Fig.~\ref{fig:Method}B. First, we generate the probability distributions $y(\vec{\theta},x,t)$ of each observable $x$ at any time $t$ for incremental variations of parameters $\theta_\mu$ and evaluate the finite derivatives corresponding to the Jacobian:
\begin{equation}
\frac{\partial y(\theta_{\mu},x,t)}{\partial\theta_{\mu}} = \frac{y(\theta_{\mu}+\Delta\theta_{\mu},x,t)-y(\theta_{\mu}-\Delta\theta_{\mu},x,t)}{2\Delta\theta_{\mu}}
\label{eq:fider}
\end{equation}
First, we generate $2N+1$ probability distributions $y(\vec{\theta}\pm\Delta\vec{\theta},x,t)$ for each observable $x$, for a model with $N$ parameters. All the probabilities used in equation~\ref{eq:fider} need to be normalized to the number of points that generated the probability distribution. Second, we calculate the eigenvalues of the FIM by summing over the entire observable landscape. Third, we track the eigenvalues of the FIM over time. In general, the eigenvalues of the FIM are logarithmically-distributed~\cite{Transtrum_Sethna_2011}. The important feature of the eigenvalues is not their absolute values but rather their relative values, which is to say that the largest eigenvalue points to the most important direction in parameter space.

When evaluating the finite derivatives in equation~\ref{eq:fider}, the choice of $\Delta\theta$ is arbitrary. In our experience, the most robust choice to avoid numerical instability and artifacts while keeping significant changes is to incrementally vary energies by $0.05$ $k_{B}T$ (leading to a change of $5\%$ for corresponding rate constants, see Appendix S$2$).
\begin{figure*}[!htb]
\centering
\includegraphics[width=0.8\textwidth]{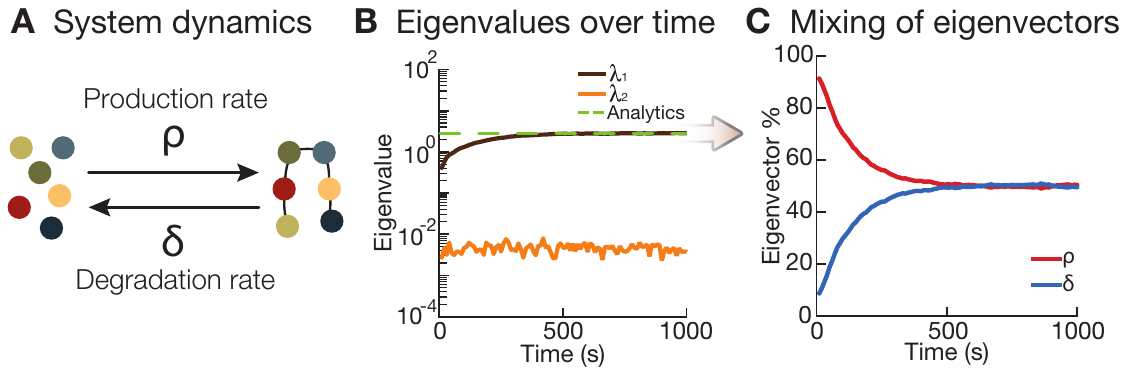}
\caption{\fontencoding{T1}\fontfamily{phv}\selectfont\textbf{Protein production and degradation.} (A) Schematic of a simple protein production-degradation system with production rate $\rho$ and degradation rate $\delta$. (B) Plot of eigenvalues over time for the protein production-degradation system. There is one dominating eigenvalue and it matches the analytical result. (C) Plot of eigenvector $\%$ from the dominating eigenvalue of panel B. The production rate $\rho$ dominates at early time points but at stationarity, the production rate and degradation rate contribute equally. Note: The eigenvector $\%$ is the absolute value of the parameter component.}
\label{fig:Protein}
\end{figure*}

\vspace{-10pt}
\section*{RESULTS}
\vspace{-10pt}
\subsection*{Test Case: One-dimensional random walk}
\vspace{-5pt}
To test our numerical PSC method, we benchmarked our algorithm by simulating a model for which an analytical solution is available. We chose the one-dimensional random walk model introduced in Machta \textit{et al.}~\cite{Machta_Sethna_2013}, which is the model used to develop the concept of PSC. The parameters of the model are the probabilities of a particle jumping to one of six neighboring sites (Fig.~\ref{fig:Method}B); the observable $x$ of the model is the position of a particle and $y(\vec{\theta},x,t)$ is the distribution of particle positions as a function of time (\emph{viz}., the particle density in a mean-field approximation when there are many particles). We simulated the random walk and plotted the eigenvalues of the FIM over time (Fig.~\ref{fig:Method}C), and our results precisely match those derived from the analytical expression (see Appendix S3). In particular, the eigenvalues start at unity; as time progresses, the distribution of eigenvalues expands, establishing a clear hierarchy of eigenvalues at later times.

As pointed out in Machta \textit{et al.}~\cite{Machta_Sethna_2013}, the first two eigenvalues can be interpreted as a drift term and a diffusion coefficient, respectively; the spreads of the eigenvalues are enough to reproduce most of the data in an effective theory (as discussed in~\cite{Machta_Sethna_2013}). We further tested the correspondence between our numerical results and the analytical theory by introducing drift into the random walk, which was not done previously. The particle density over time is shown in Fig.~S\ref{fig:RW_combine}A. The eigenvalues of the FIM over time for the perturbed random walk are shown in Fig.~S\ref{fig:RW_combine}B. The result is similar to uniform diffusion and we are able to show that the eigenvalues are defined by the probabilities of particles jumping to neighboring sites. Importantly, our numerical results precisely match the analytical solutions we derived for a random walk with drift and a uniform random walk with different numbers of parameters (see Fig.~S\ref{fig:RW_combine}C, D respectively). Thus, our numerical PSC method successfully compressed this classic physical system.

\vspace{-10pt}
\subsection*{Test case: A simple protein production and degradation system}
\vspace{-5pt}
Having benchmarked our algorithm against the random walk model, we next wondered how our numerical PSC method would handle a model where the distribution of an observable is determined by a combination of the initial parameters. Such situations will arise in most if not all real computational models. Therefore, we applied our numerical PSC method to a textbook biophysical model of protein production and degradation. The model has only two parameters, the production rate $\rho$ and the degradation rate $\delta$ (see Fig.~\ref{fig:Protein}A). The observable $x$ of the model is the number of proteins in the system at any given time. Importantly, The stationary distribution $y(\vec{\theta},x)$ of protein number is a Poisson distribution of the parameter combination $\rho/\delta$ (representing the expectation value for the number of proteins)~\cite{Ross_2010}. Using this stationary distribution, we can analytically solve for the dominating eigenvalue of the corresponding FIM in the continuous limit:
\begin{equation}
\lambda_{1} \simeq \frac{1}{2}\sqrt{\frac{\rho}{\delta\pi}}
\label{eq:protein}
\end{equation}

\noindent The derivation of the eigenvalues and the expression for equation~\ref{eq:protein} can be found in Appendix S$4$. In the continuous limit, the second eigenvalue of the system goes to $0$, but not in a discrete simulation (see Appendix S$4$ and Fig.~S\ref{fig:Protein_combine}B).

\noindent Starting from an initial condition with no proteins, we simulated this process using the Gillespie algorithm~\cite{Gillespie_1977} and computed the eigenvalues of the system over time (Fig.~\ref{fig:Protein}B). One eigenvalue is always over 2 orders of magnitude larger than the other, indicating that the system is governed by one effective parameter, which is to say that there is only one relevant direction in parameter space that determines the model's output. Looking at the relative contribution of the eigenvector components of the dominating eigenvalue in Fig.~\ref{fig:Protein}C, we can see that during the early stage of the system, the production rate $\rho$ dominates, corresponding to the net production of proteins from the initial condition. The system then reaches stationarity, at which point the eigenvector components of the dominating eigenvalue are a mix of the the production rate $\rho$ and degradation rate $\delta$, with equal contributions as expected from our derivation (Fig.~\ref{fig:Protein}C). We checked that our method recovers the analytical result of equation \ref{eq:protein} (in the asymptotic limit) for different ratios of production rate over degradation rate (see Fig.~\ref{fig:Protein}B, S\ref{fig:Protein_combine}A). Thus, our numerical PSC method is able to compress out irrelevant directions and extract the effective parameter defining the distribution of protein number (here, a Poisson distribution).

\vspace{-10pt}
\subsection*{Microtubule dynamics: a complex biological system}
\vspace{-5pt}
Having fully characterized our method, next, we applied it to a biological system that cannot be solved analytically, namely the dynamic instability of microtubules~\cite{Mitchison_Kirschner_1984}. Microtubules are polymers of \textalpha\textbeta-tubulin and dynamic instability is the non-equilibrium behavior in which the polymers stochastically switch between periods of growth and shrinkage. This complex, non-equilibrium phenomenon was first simulated numerically in the 1980s~\cite{Chen_Hill_1985,Peter_Martin_1989} and has remained a subject of considerable interest for computational biologists, who have developed increasingly sophisticated models~\cite{VanBuren_Odde_2005,Castle_Odde_2017,Zakharov_Grishchuk_2015,McIntosh_Nikita_2018}. The long-term goal of these collective efforts is to develop a powerfully-predictive yet minimal model that can be used to explain microtubule physiology. Our numerical PSC method has the power to determine whether existing models have an underlying low-dimensional structure.

\begin{figure}[H]
\centering
\includegraphics[width=\columnwidth]{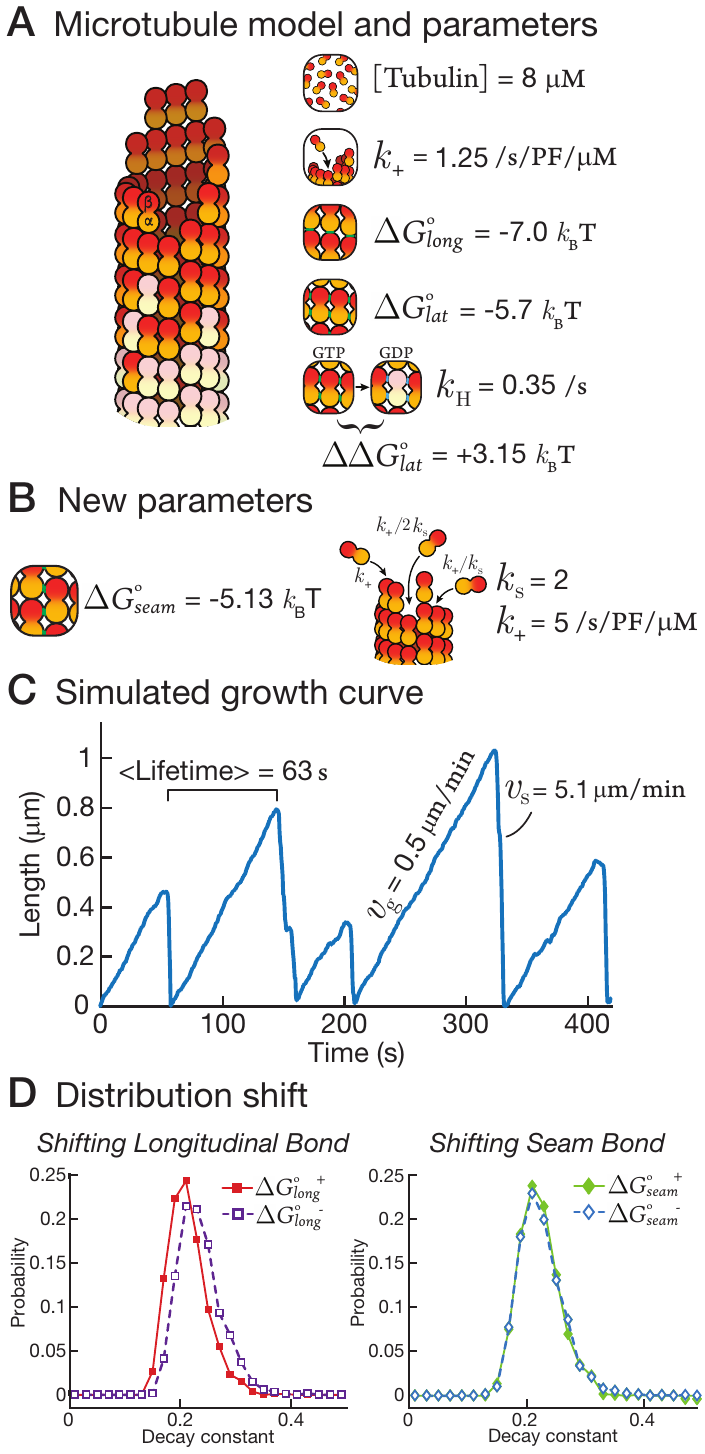}
\caption{\fontencoding{T1}\fontfamily{phv}\selectfont\textbf{Microtubule dynamics.} (A) The base model of our microtubule simulation following VaBurren \textit{et al.} (B) Two new adjustable parameters were introduced to the base microtubule model, the weaker lateral bond at the seam $\Delta G_{\mathit{seam}}^{o}$ and the neighboring penalty $k_{s}$, which decreases the on rate for tip positions that have neighboring tubulins. (C) Plot of length versus time from our microtubule simulation. (D) (\textit{Left}) Plot of the decay constant distributions of two $\Delta G_{long}^{o}$. (\textit{Right}) Plot of the decay constant distributions of two lateral bonds at the seam $\Delta G_{seam}^{o}$.}
\label{fig:MT_system}
\end{figure}

Our starting model is based on VanBuren \emph{et al.} 2002 (see Fig.~\ref{fig:MT_system}A)~\cite{VanBuren_Odde_2002}; a similar model is used by Ayaz \textit{et al.} 2014~\cite{Ayaz_Luke_2014}. We chose this model because it's a classic and because understanding its underlying dimensionality will inform ongoing modeling work on microtubules. Briefly, tubulin subunits associate head-to-tail to create protofilaments (pfs), forming longitudinal bonds described by an energy parameter $\Delta G_{long}^{o}$. In our model, 13 pfs are connected by lateral bonds between adjacent subunits with an energy parameter $\Delta G_{lat}^{o}$ ~\cite{Ledbetter_Porter_1964}. \textbeta-tubulin forms lateral bonds with other \textbeta-tubulins (and \textalpha with \textalpha) except at a single discontinuity in the lattice known as the ``seam'', where \textbeta-tubulin binds to an \textalpha-tubulin (see below). The rate at which tubulin binds to the microtubule ends is described by an association rate constant, $k_{+}$. Because tubulin is a GTPase, these incoming tubulin subunits contain GTP in the \textbeta-tubulin nucleotide pocket. This GTP becomes hydrolyzed after (1) the subunit incorporates into the polymer and (2) another GTP-tubulin binds on top of it, contributing catalytic residues that complete the nucleotide pocket~\cite{Lowe_Nogales_2001}. The rate of GTP hydrolysis is described by a rate constant parameter $k_{H}$. GTP hydrolysis and phosphate release converts GTP-tubulin to GDP-tubulin and weakens the bonds between tubulin subunits in the polymer~\cite{Mandelkow_Milligan_1991,Maurer_Surrey_2012}. Following VanBuren, this weakening of energies is described by an energy parameter, $\Delta\Delta G_{lat}^{o}$, which is assigned to the lateral bonds of the new GDP-tubulin subunit. Using parameter values similar to Castle \textit{et al.} 2017 (see Fig.~\ref{fig:MT_system}B), our simulation produces microtubule growth curves that correspond reasonably with in-house experimental data generated by \emph{in vitro} reconstitution experiments at $8$ $nm$ tubulin~\cite{Chaaban_Brouhard_2018} (see Fig.~\ref{fig:MT_system}C). More specifically, microtubules grow as long as their ends are protected by a ``cap'' of GTP-tubulin~\cite{Carlier_Pantaloni_1981}. If this GTP cap is ``lost'', the polymer switches to rapid shrinkage in an event known as a ``catastrophe'', the hallmark of dynamic instability~\cite{Mitchison_Kirschner_1984}.

There are many subtleties and caveats to models of dynamic instability. For example, which bonds are weakened by GTP hydrolysis is not well established~\cite{Zhang_Nogales_2015,Manka_Moores_2018}, and the transition from GTP-tubulin to GDP-tubulin may have substeps~\cite{Manka_Moores_2018}. These subtleties are discussed in Appendix S$5$. We used the direct method of the Gillespie algorithm~\cite{Gillespie_1977}, which is a different implementation than the one in VanBuren \textit{et al.}~\cite{VanBuren_Odde_2002} and Ayaz \textit{et al.}~\cite{Ayaz_Luke_2014}. In order to validate our Gillespie algorithm, we used the parameters found in Ayaz \textit{et al.} and confirmed that our simulation produces identical results. The details of our simulation method and the benchmarking of our algorithm against published data can be found in Appendix S$5$ and Fig.~S\ref{fig:MT_Benchmark_combine}.

\begin{figure*}[!htb]
  \begin{minipage}[c]{0.67\textwidth}
    \centering
    \includegraphics[width=\textwidth]{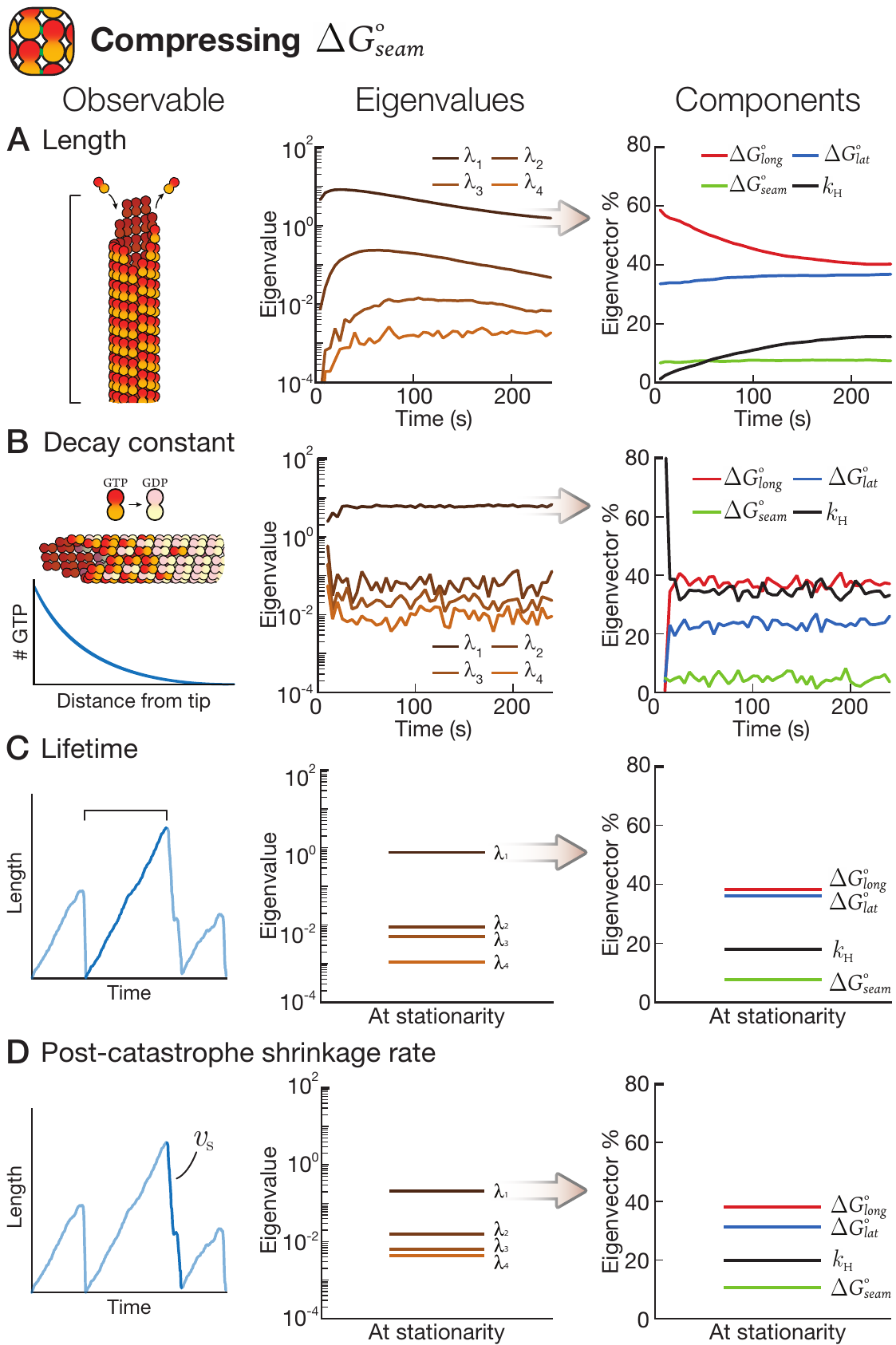}
  \end{minipage}\hfill
  \begin{minipage}[c]{0.3\textwidth}
  \vspace{-340pt}
    \caption{\fontencoding{T1}\fontfamily{phv}\selectfont\textbf{Numerical PSC for the base microtubule model plus $\Delta G_{seam}^{o}$.} Eigenvalues and eigenvector components ($\%$) for four observables: (A) Length (the exact number of dimers in each microtubule), (B) Decay constant of GTP from the tip, (C) Average lifetime of the microtubule, (D) Post-catastrophe shrinkage rate. Note: The eigenvector $\%$ is the absolute value of parameter component.}
    \label{fig:MT_normal}
  \end{minipage}
\end{figure*}

To our minimal model, we sequentially added parameters that are motivated by recent discoveries and controversies concerning the structure of the microtubule lattice and the microtubule end.  We then ``compressed'' this complex model to see whether the new parameters are essential (\emph{viz}., become components of the dominating eigenvalues of the FIM) and whether new parameters increase the underlying dimensionality of the model. Following VanBuren \textit{et al.}, we varied 3 core parameters ($\Delta G_{long}^{o}$, $\Delta G_{lat}^{o}$, and $k_{H}$) and treated the others as non-adjustable ($k_{+}$, $\Delta\Delta G_{lat}^{o}$). In order to compress our microtubule model, we measured four independent observables of the simulations that correspond to experimental data~\cite{Chaaban_Brouhard_2018}. Two observables can be tracked continuously over the time course of the simulation: (1) the length of microtubule (Fig.~\ref{fig:MT_normal}A) and (2) the decay constant that
\noindent describes the conversion of GTP-tubulin into GDP-tubulin (``GTP cap size'', Fig.~\ref{fig:MT_normal}B)~\cite{Bieling_Thomas_2007}. The second column of Fig.~\ref{fig:MT_normal}A and B shows the eigenvalues over time for these observables. The other two observables are the distribution of microtubule lifetimes (Fig.~\ref{fig:MT_normal}C) and the post-catastrophe shrinkage rate (Fig.~\ref{fig:MT_normal}D). These observables are not tracked continuously because they require post-simulation analysis to determine when catastrophes occurred (see Appendix S6). The second column of Fig.~\ref{fig:MT_normal}C and D shows the eigenvalues for these observables at the conclusion of the simulation, when the distributions have reached stationarity. This framework allowed us to apply our numerical PSC method to this model of microtubule dynamic instability and test the importance of new parameters.

\vspace{-10pt}
\subsubsection*{The ``seam'' parameter can be compressed away}
\vspace{-5pt}
We first added a parameter to account for the atypical lateral bonds that form at the ``seam'' in the lattice, $\Delta G_{\mathit{seam}}^{o}$. These atypical bonds are presumably weaker than other lateral bonds ($\Delta G_{\mathit{seam}}^{o} < \Delta G_{lat}^{o}$), a presumption that is supported by molecular dynamics simulations~\cite{Sept_McCammon_2003} and recent atomic resolution cryo-electron microscopy, in which the lateral bonds at the seam are slightly ``open'' relative to other lateral bonds~\cite{Zhang_Nogales_2015,Zhang_Nogales_2018}. The apparent weakness of the seam has lead to the proposal that microtubules first crack open at the seam when a catastrophe begins. We wanted to test this proposal by explicitly including the seam in our simple model.

We generated the FIM for a simulation of the VanBuren \textit{et al.} model with an explicit seam: $\Delta G_{\mathit{seam}}^{o} = 0.9\cdot\Delta G_{lat}^{o}$ based on~\cite{Sept_McCammon_2003}. For all four observables, one eigenvalue strongly dominates (note the log-scale for eigenvalues). This dominance implies that the distribution of each observable is determined by a single effective parameter.  This result is not obvious: one expects that the mean and the variance of any given distribution are described by independent parameters, as was the case for the random walk~\cite{Machta_Sethna_2013}. Rather, microtubule observables are similar to the number of proteins in the protein production/degradation model, where both the mean and variance of the distributions are determined by a single effective parameter.

As for the protein production/degradation case, the single effective parameter determining the distribution of each observable is \textit{a priori} a complex function of the initial parameters. The relative influence of each initial parameter is given by the eigenvector components of the dominant eigenvalue (see column three of Fig.~\ref{fig:MT_normal}A-D). Importantly, we also can see which parameters are \textit{not} important for a given observable, because these initial parameters will be insignificant components of the dominating eigenvalue.
The important components for microtubule length are the longitudinal bond, $\Delta G_{long}^{o}$, followed closely at later times by the lateral bond, $\Delta G_{lat}^{o}$. This is not surprising, considering that these bond energies are what drive polymerization. The important components for the decay constant are more interesting: in addition to the obvious parameter of the GTP hydrolysis rate constant, $k_{H}$, the decay constant is also determined by $\Delta G_{long}^{o}$ and $\Delta G_{lat}^{o}$. A simple interpretation of this result is that a microtubule that forms stronger bonds (and hence grows faster) will have a larger GTP cap. Consistently, microtubules that grow faster have larger GTP caps when End Binding proteins are used as reporters of GTP cap size~\cite{Bieling_Thomas_2007}. The lifetime distribution and the post-catastrophe shrinkage rates are similarly complex, depending on both bond energies and $k_{H}$. Interestingly, $k_{H}$ makes a relatively minor contribution to the microtubule lifetime distribution. This result implies that the best way to avoid a catastrophe might not be to hydrolyze GTP slightly slower but rather to form slightly stronger bonds and hence grow faster (see Discussion).
\begin{figure*}[!htb]
  \begin{minipage}[c]{0.67\textwidth}
    \centering
    \includegraphics[width=\textwidth]{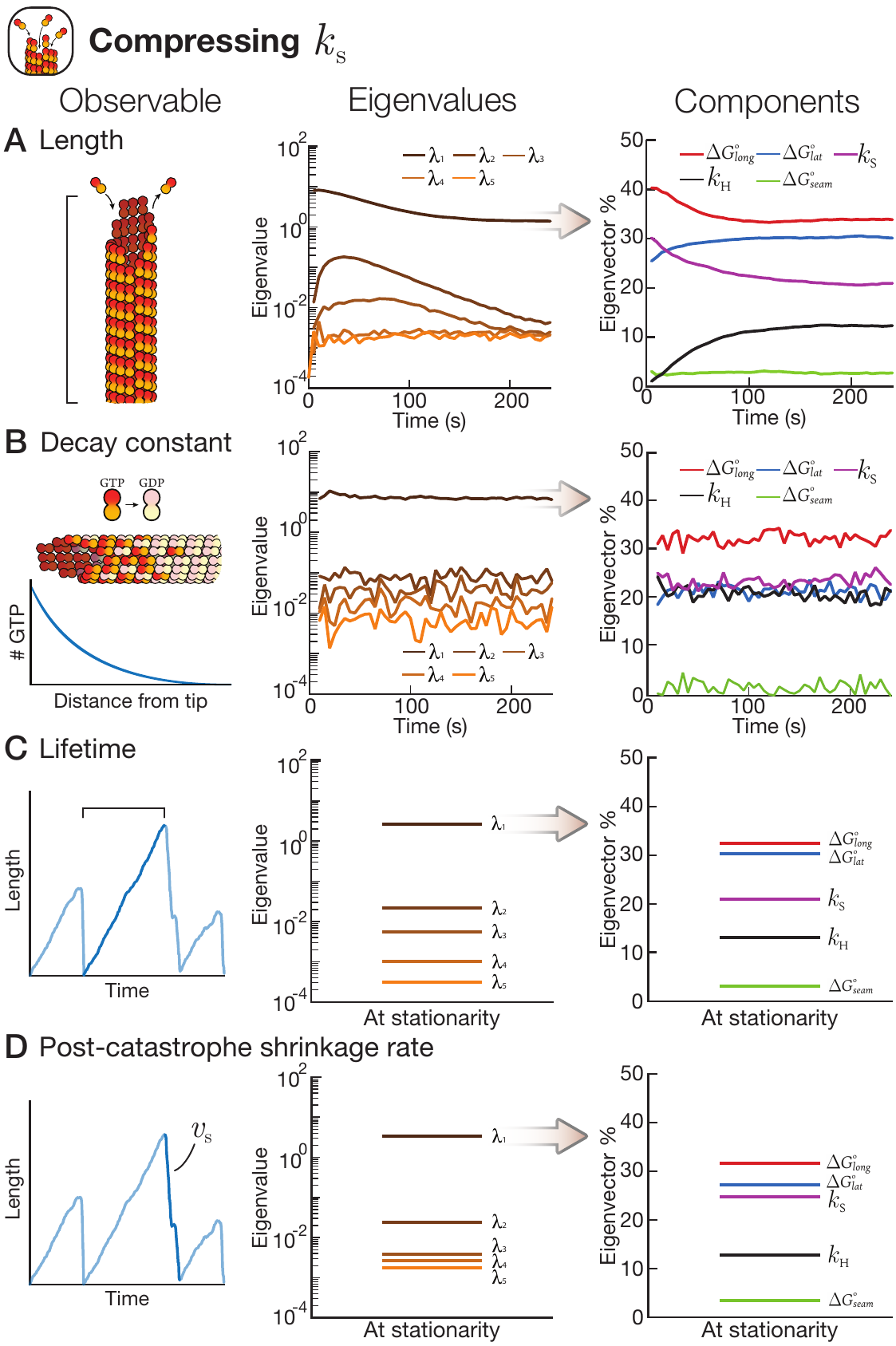}
  \end{minipage}\hfill
  \begin{minipage}[c]{0.3\textwidth}
  \vspace{-320pt}
    \caption{\fontencoding{T1}\fontfamily{phv}\selectfont\textbf{Numerical PSC for the base microtubule model plus $\Delta G_{seam}^{o}$ and $k_{s}$.} Eigenvalues and eigenvector components ($\%$) for four observables: (A) Length (the exact number of dimers in each microtubule), (B) Decay constant of GTP from the tip, (C) Average lifetime of the microtubule, (D) Post-catastrophe shrinkage rate. Note: The eigenvector $\%$ is the absolute value of parameter component.}
    \label{fig:MT_diffrate}
  \end{minipage}
\end{figure*}

But for all four observables, $\Delta G_{\mathit{seam}}^{o}$ was an insignificant component of the dominant eigenvalue. In other words, the ``seam'' parameter has been compressed away. We can confirm the insignificance of the seam directly by plotting the model's output for two different parameter values. Fig.~\ref{fig:MT_system}D (\textit{left}) shows the distribution of decay constants at two values of $\Delta G_{long}^{o} \pm 0.1\ k_{B}T$; the difference in the distributions is clear. In contrast, Fig.~\ref{fig:MT_system}D (\textit{right}) shows the decay constant distribution at two values of $\Delta G_{\mathit{seam}}^{o} \pm 0.1\ k_{B}T$; the distributions are indistinguishable, indicating that changes in $\Delta G_{\mathit{seam}}^{o}$ did not impact the model's outputs. Of course, we cannot rule out that $\Delta G_{\mathit{seam}}^{o}$ will have a significant impact on an observable not measured here. Nevertheless, while the seam may have interesting structural characteristics~\cite{Zhang_Nogales_2015,Zhang_Nogales_2018}, it can be safely ignored in simple models of microtubule {dynamics}.

\vspace{-10pt}
\subsubsection*{Accounting for end structure is essential}
\vspace{-5pt}
Next, we added a parameter to account for the tapering of pfs found at microtubule ends~\cite{Chretien_Karsenti_1995}. Because of tapering, some incoming dimers will bind to sites with lateral neighbors and some will not. Recent work from Castle \textit{et al.}~\cite{Castle_Odde_2013} used Brownian dynamics simulations to show that lateral neighbors block off some of the diffusional paths that lead a tubulin dimer to its binding site. The blocking of these paths creates a ``penalty'' for the association rate constant $k_{+}$. As such, we introduce $k_{s}$, a parameter that reduces the association rate constant $k_{+}$ based on the number of lateral neighbors at each binding site~\cite{Gardner_Odde_2011,Castle_Odde_2017}. Our value for $k_{s}$ is roughly equivalent to the penalties produced by Castle \textit{et al.}'s Brownian dynamics simulations. We simulated microtubules and calculated the FIM for the system, as above. Fig.~\ref{fig:MT_diffrate}A-D follow Fig.~\ref{fig:MT_normal}, showing our four observables, the eigenvalues, and eigenvector components.

Despite the introduction of a new parameter, one eigenvalue strongly dominates each observable (Fig.~\ref{fig:MT_diffrate}). Thus, the introduction of a new parameter did not add an additional dominating eigenvalue, meaning that the underlying dimensionality of the system did not change. This result is also not obvious. Adding a new parameter like $k_{s}$ might have added a new relevant direction in parameter space, but here it did not. Each observable is still governed by a single effective parameter.

Although the addition of $k_{s}$ did not add a new effective parameter, $k_{s}$ does appear in the eigenvector components of the dominant eigenvalues for all 4 observables (see column 3 of Fig.~\ref{fig:MT_diffrate}). Indeed, $k_{s}$ makes a contribution equal to or greater than $k_{H}$. This result indicates that accounting for $k_{s}$, and more generally for the complexity of the structure of microtubule ends, is essential even in simple models of microtubule dynamics that track only a few observables.

\vspace{-10pt}
\subsubsection*{Estimating the dimensionality of the system}
\vspace{-5pt}
As shown above, the distribution of each observable (e.g., microtubule lifetime) can be described by a single effective parameter, given by the eigenvector components of the dominating eigenvalue. But are these effective parameters the same for all observables or are there four orthogonal effective parameters? Looking at the eigenvector components for the length, the microtubule lifetime distribution, and the post-catastrophe shrinkage rate, it's clear that $\Delta G_{long}^{o}$, $\Delta G_{lat}^{o}$, and $k_{s}$ are important components of the effective parameter for all three of these observables. So perhaps the effective parameter here is the same. In contrast, the decay constant has a significant eigenvector component from the hydrolysis rate constant $k_{H}$, indicating that the effective parameter determining the decay constant might differ from the others. So what is the true dimensionality of our microtubule model?
\begin{figure}[H]
\centering
\includegraphics[width=0.5\columnwidth]{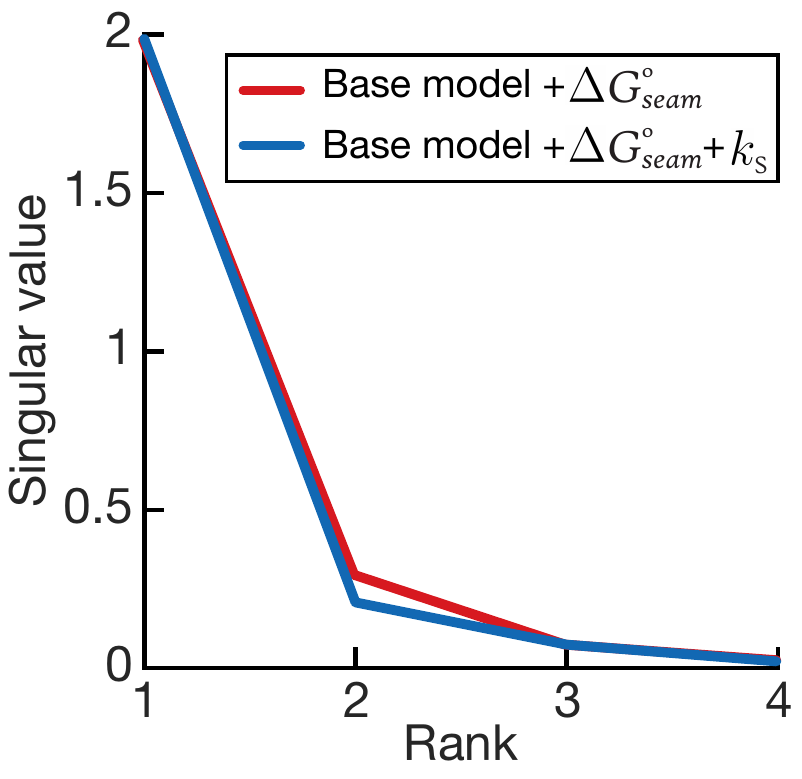}
\caption{\fontencoding{T1}\fontfamily{phv}\selectfont\textbf{Identifying the dimensionality of the models.} Plot of the singular values for the base model plus $\Delta G_{seam}^{o}$ and the base model plus $\Delta G_{seam}^{o}$ and $k_{s}$. Both systems are two-dimensional.}
\label{fig:SVD_compare}
\end{figure}
In order to estimate the dimensionality of the model, we need to consider all observables as a whole. Intuitively, the dimensionality of the system is defined by the dominating eigenvalues. To perform a more rigorous estimation of dimensionality for each model, we computed the singular value decomposition (SVD) on the eigenvectors of the four dominating eigenvalues corresponding to the four observables~\cite{Rocha_2003}. We performed the SVD analysis for both our base model with $\Delta G_{\mathit{seam}}^{o}$ (Fig.~\ref{fig:SVD_compare}, \emph{red}) and our base model with $\Delta G_{\mathit{seam}}^{o}$ and $k_{s}$ (Fig.~\ref{fig:SVD_compare}, \emph{blue}). Interestingly, in both cases SVD revealed a sloppy distribution of singular values, with one dominating singular value, a second singular value roughly one order of magnitude smaller, and two even smaller singular values one order of magnitude below. The presence of two large singular values means that two effective parameters are enough to fit the data with close to a 99\% precision. The vector components for the base model with $\Delta G_{\mathit{seam}}^{o}$ and base model with $\Delta G_{\mathit{seam}}^{o}$ and $k_{s}$ are shown in Fig.~S\ref{fig:SVD_combine}A and Fig.~S\ref{fig:SVD_combine}B respectively.

This analysis demonstrates rigorously that the full dimensionality of our model is roughly equal to two in both cases. The first effective parameter drives the length, the microtubule lifetime distribution, and the post catastrophe shrinkage rate, with extended contributions from lattice energies and a smaller contribution from the hydrolysis rate constant. We call this parameter the ``effective polymerization'' parameter. The second effective parameter drives the decay constant and is a combination of the $k_{H}$ and $\Delta G_{lat}^{o}$. We call this parameter the ``effective hydrolysis'' parameter. It is remarkable that dynamic instability can be compressed into a two parameter system. Interestingly, the addition of $k_{s}$ did not change the underlying dimensionality of the microtubule model. Rather, $k_{s}$ simply became a significant component of the effective polymerization parameter.

\vspace{-10pt}
\section*{DISCUSSION}
\vspace{-5pt}
As biophysicists, we want to capture the complexity of biology in the simplest possible terms, even if those terms are themselves quite complex. Our work has demonstrated the power of numerical PSC as a method for identifying the essential parameters and low-dimensional structure of complex models. We first validated our method against two analytically-solvable models and then applied it to a well-known computational model of microtubule dynamic instability. Thus, our method opens the door to the simplification of any computational model in biology.

Our application of numerical PSC to a simple model of microtubule dynamic instability provided four direct insights into microtubule growth and catastrophe. First, we were surprised to find that $k_{H}$, the hydrolysis rate constant, had a lesser influence on the lifetime distribution relative to $\Delta G_{long}^{o}$ and  $\Delta G_{lat}^{o}$. This suggests that catastrophe might be more efficiently prevented by making stronger bonds rather than by slowing down hydrolysis. Our interpretation is that stronger bonds help prevent pfs from losing their terminal GTP-tubulin dimers, which would cause the pf to become ``fully uncapped''. Bowne-Anderson \textit{et al.}\ argued that uncapping of pfs are the irreversible events that lead to catastrophe~\cite{Bowne-Anderson_Howard_2013}. Similarly, poisoning of pf ends with the drug eribulin has a very strong effect on catastrophe frequency~\cite{Doodhi_Akhmanova_2016}. Therefore, our results showing the importance of $\Delta G_{long}^{o}$ and  $\Delta G_{lat}^{o}$ are consistent with the emerging concept that ``pf destabilization'' is a root cause of catastrophe.

The second conclusion is that $k_{s}$ contributes significantly to the effective parameters controlling all four observables. $k_{s}$ is important because it significantly changes the rate at which longitudinal bonds are formed. Parameters like $k_{s}$ were absent from early models of microtubule dynamics~\cite{VanBuren_Odde_2002} before the importance of tapering of microtubule ends was clear~\cite{Coombes_Gardner_2013}. Additional parameters related to end structure, such as the lateral and longitudinal curvature of tubulin dimers as they enter the lattice~\cite{Brouhard_Rice_2018}, are sure to determine microtubule behavior.

Third, the insignificance of $\Delta G_{seam}^{o}$ demonstrates that the seam's slight weakness relative to other lateral bonds is unlikely to determine the fate of a microtubule in terms of its growth rate or lifetime. Interestingly, increasing the number of seams in a microtubule experimentally does not change its measured mechanical properties~\cite{Harris_Hawkins_2018}. Therefore, although the seam is an interesting structural feature of the microtubule~\cite{Zhang_Nogales_2018}, accounting for it is not necessary in simple computational models, at least as far as our observables are concerned. We look forward to expanding our analysis to more complex models and larger sets of observables.

Lastly, our analysis of dimensionality revealed that almost all data simulated here can be described with only two parameters, corresponding to an effective polymerization parameter and an effective hydrolysis parameter. Computation of such effective parameters is made rigorous and possible by the use of our approach. The addition of new parameters (such as $k_s$) might contribute to the effective parameters of a model without changing dimensionality. Interestingly, our approach could help experimentalists identify the types of data that are necessary to define the effective parameters. In the microtubule case, we see that the decay constant directly identifies the effective hydrolysis parameter.

Our ability to distinguish between models in science is always limited by the availability of hard data (except in string theory). Which parameters of a model are ``stiff'' and which are ``sloppy'' depends critically on the observables that the model attempts to reproduce. In biophysics, the rigor of physical modeling collides against the complexity of biological interactions. A coupling of theory and experiment is necessary to disentangle this complexity. PSC tightens this coupling by improving the interpretability of models, which in turn identifies the key experiments that drive theory forward.

\vspace{-10pt}
\section*{ACKNOWLEDGEMENTS}
\vspace{-5pt}
We thank Sami Chaaban, Claire Edrington, Dr.~Hadrien Mary, Félix Proulx-Giraldeau, Thomas Rademaker, Laurent Jutras-Dubé and Dr.~Adrien Henry for feedback on this project and comments on the manuscript. C.T.H. acknowledges support from the Milton Leung fund in McGill Physics and from Fonds de recherche du Quebec - Nature et technologies (FRQNT) Bourse. G.J.B. is supported by the Canadian Institutes of Health Research: Operating Grant $\#137055$, Operating Grant $\#$PJT-$148702$ and the Natural Sciences and Engineering Research Council of Canada: Discovery Grant $\#372593$. P.F. is supported by a Simons Foundation fellowship in Mathematical Modelling of Biological Systems and NSERC Discovery Grant $2016$-$06501$. Lastly this work was supported by a FRQNT Projet de Recherche en Equipe $\#$FRQ-NT$191128$.

\newpage
\printbibliography[title={REFERENCES}]

\newpage
\setcounter{equation}{0}
\setcounter{figure}{0}
\captionsetup{labelformat=empty}
\begin{refsection}
\section*{Appendix S1: Derivation of the Fisher Information Matrix (FIM) expression following Machta \textit{et al.} 2011~\cite{Transtrum_Sethna_2011}}
\vspace{-5pt}
The FIM expression presented in the main text is a way to estimate how model parameters can be fitted to data, and by extension how models with different parameters are distinguishable. We will use the FIM in the latter sense, and in the following recall its connection to data fitting.

Assume we have a mathematical model of a biological systems, with parameters $\vec{\theta}$, giving the probability $y\left(\vec{\theta},x_i\right)$ of observable $x$ to take the value $x=x_i$.
We call $d_i$ the experimentally measured  probability of value $x_i$ and assume that:
\begin{equation}
d_{i} = y\left(\vec{\theta},x_{i}\right) + \sigma_i r_{i}
\end{equation}
where we assume $r_{i}$ to be a random Gaussian noise of mean $0$ and variance $1$, and $\sigma_i$ a local variance so that:
\begin{equation}
r_{i}\left(\vec{\theta}\right) = \frac{d_{i} - y\left(\vec{\theta},x_{i}\right)}{\sigma_{i}}
\label{eq:residue}
\end{equation}

Assuming all $r_i$s are independent, the total probability $P(\vec{r},\vec{\theta})$ of experimentally observing the values $\vec{d}=\{d_{i}\}$ given the residuals $\vec{r}=\{r_{i}\}$,is thus :
\begin{equation}
P\left(\vec{r},\vec{\theta}\right) = \frac{1}{(2\pi)^{M/2}}\exp\left(-\frac{1}{2}\sum_{i=1}^Mr_{i}\left(\vec{\theta}\right)^2\right)
\end{equation}
where $M$ is the number of points where we try to fit data.

\noindent The Fisher Information Matrix (FIM) defines the amount of information that the residuals $r_i$  contain on parameters. Intuitively, the FIM tells us about the distinguishability of two parameter sets given the data. It is given by:
\begin{equation}
\begin{aligned}
I_{\mu,\nu} &= \left\langle -\frac{\partial^2 \log P\left(\vec{r},\vec{\theta}\right)}{\partial\theta_{\mu}\partial\theta_{\nu}}\right\rangle \\
            &=-\int d \vec{r} P\left(\vec{r},\vec{\theta}\right) \frac{\partial^2 \log P\left(\vec{\theta},r_i\right)}{\partial\theta_{\mu}\partial\theta_{\nu}}\\
\end{aligned}
\end{equation}

Since $r_i$ are Gaussian distributed, one can explicitly perform the computation of the FIM, which can then be interpreted as a metric  $g_{\mu,\nu}$ quantifying the ability to distinguish between different parameter sets. Following Machta \textit{et al.} we have:

\begin{equation}
\begin{aligned}
g_{\mu,\nu} &=\left\langle-\frac{\partial^2\log P\left(\vec{\theta},\xi\right)}{\partial\theta_{\mu}\partial\theta_{\nu}}\right\rangle = \left\langle\frac{\partial^2\sum_{i}\frac{1}{2}r_{i}^2}{\partial\theta_{\mu}\partial\theta_{\nu}}\right\rangle\\
			&=\sum_{i}\left\langle r_{i}\frac{\partial^2 r_{i}}{\partial\theta_{\mu}\partial\theta_{\nu}}+\frac{\partial r_{i}}{\partial\theta_{\mu}}\frac{\partial r_{i}}{\partial\theta_{\nu}}\right\rangle\\
            &= \sum_{i}\left\langle r_{i}\frac{\partial^2 r_{i}}{\partial\theta_{\mu}\partial\theta_{\nu}}\right\rangle +
            \sum_{i}\left\langle\frac{\partial r_{i}}{\partial\theta_{\mu}}\frac{\partial r_{i}}{\partial\theta_{\nu}}\right\rangle\\
\end{aligned}
\label{eq:FIM_res}
\end{equation}

\noindent We substitute equation~\ref{eq:residue} into equation~\ref{eq:FIM_res}. The first sum cancels out because $r_i$ is independent of $y\left(\vec{\theta},x_{i}\right)$, $d_i$ is independent of $\vec{\theta}$ and the expectation value of the residue  $r_i$ itself is zero. The second term becomes fully deterministic since with the same assumptions $\frac{\partial r_{i}}{\partial\theta_{\nu}}= \frac{\partial y\left(\vec{\theta},x_{i}\right)}{\partial \theta_{\nu}}$. We arrive at the final expression for the Fisher Information Matrix of the model:
\begin{equation}
g_{\mu,\nu} = \sum_{x} \frac{\partial y\left(\vec{\theta},x\right)}{\partial \theta_{\mu}} \frac{\partial y\left(\vec{\theta},x\right)}{\partial \theta_{\nu}}
\end{equation}
where we assume all $\sigma_i$ to be equal (and rescaled to $1$).
\noindent The remarkable result is that this metric, while initially computed by averaging over experiments, is a pure function of the model $y$, and as a consequence, can be used independently of actual experiments to estimate in a deterministic way models distinguishability.

\noindent Biological models mix rates and energies, so the parameters are potentially of very different nature. Mixing units might yield purely dimensional effects in the analysis of important directions in parameter space. Energy is the most fundamental quantity and kinetic rates are exponentially related to the rescaled energy (in unit of $k_{B}T$). We take the derivatives with respect to the $\log$ of the parameter (\textit{i.e.} $\frac{\partial y\left(\vec{\theta},x_{i}\right)}{\partial\log\theta_{mu}} = \frac{\partial y\left(\vec{\theta},x_{i}\right)}{\partial\theta_{mu}}\theta_{mu}$) to express every parameters in the Fisher Information Matrix with the same effective unit. Therefore, the final expression becomes:
\begin{equation}
g_{\mu,\nu} = \sum_{x} \frac{\partial y\left(\vec{\theta},x\right)}{\partial \theta_{\mu}} \frac{\partial y\left(\vec{\theta},x\right)}{\partial \theta_{\nu}}\theta_{\mu}^{\alpha_\mu}\theta_{\nu}^{\alpha_\nu}
\label{eq:FIMexp}
\end{equation}

where $\alpha_\mu=0$ if $\theta_\mu$ is an energy and $\alpha_\mu=1$ if $\theta_\mu$ is a kinetic rate.

\vspace{-10pt}
\section*{Appendix S2: Conversion for energy $k_{B}T$ to rates $s^{-1}$ in the microtubule model}
\vspace{-5pt}
\noindent As shown in Appendix S1, biological models mix energies and rate constants as parameters. While energies are the most fundamental quantities for our Fisher Information Matrix computation, computing derivatives by incremental changes of energies might be challenging since small changes of energies could potentially give big changes of rates and thus of probability distributions. For numerical computations, we vary rate constants by a small increment \textalpha, and shift energies such as $\Delta G_{long}^{o}$, $\Delta G_{lat}^{o}$ and $\Delta G_{seam}^{o}$ logarithmically. In order to ensure a smoother change of probability distribution, we need to come up with a conversion between changing bond energy and the change in rate.

\noindent As an example, the off rate of a dimer for the microtubule model is given as:
\begin{equation}
k_{off} = \frac{k^{+}}{e^{-\Delta G_{tot}^{o}}}
\end{equation}
where $k^{+}$ is the apparent on rate and $\Delta G^{o}_{tot}$ is the total bond energy associated with the particular dimer.

\begin{figure}[H]
\centering
\includegraphics[width=0.5\columnwidth]{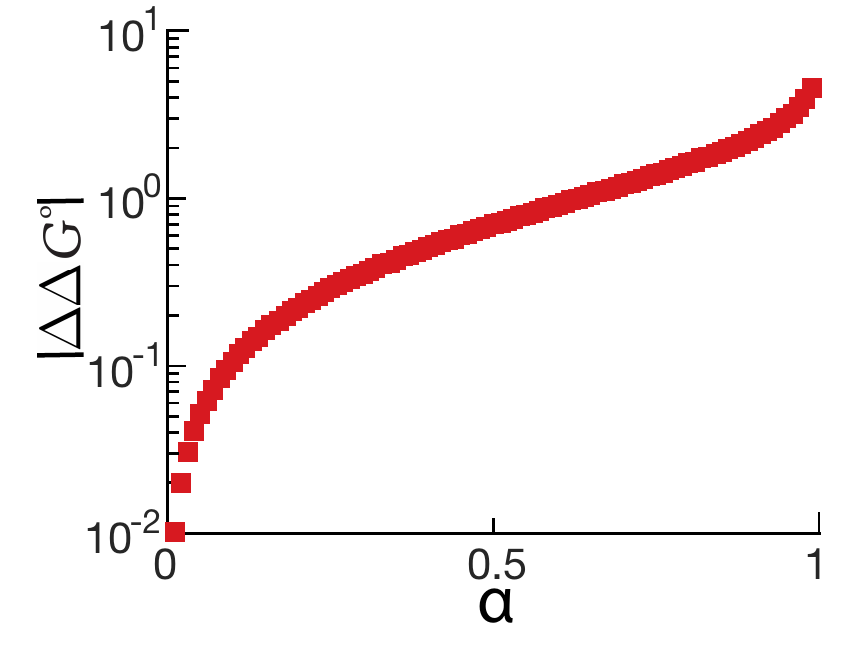}
\caption{\fontencoding{T1}\fontfamily{phv}\selectfont{\bf Figure S\ref*{fig:MT_Rateconversion}.} The conversion plot for various percentage change for parameters that have a unit of $k_{B}T$. The conversion rate is universal and it does not depend on the value of the parameter itself. Note: a $5\%$ change in parameter value is about $0.05$ $k_{B}T$.}
\label{fig:MT_Rateconversion}
\end{figure}

\noindent Let us define the original off rate $k_{ori}^{-}$ and the new off rate after changing the bond strength $k_{new}^{-}$.. Assume the difference factor is defined as the following:
\begin{equation}
\alpha = \frac{k_{ori}^{-}-k_{new}^{-}}{k_{ori}^{-}}
\end{equation}
which implies that
\begin{equation}
\begin{aligned}
k_{new}^{-}  &= k_{ori}^{-}(1-\alpha)\\
\Rightarrow \frac{k^{+}}{e^{-\Delta G_{new}^{o}}}  &= \frac{k^{+}}{e^{-\Delta G_{ori}^{o}}}(1-\alpha)\\
\Rightarrow e^{\Delta G_{new}^{o}}  &= e^{\Delta G_{ori}^{o}}(1-\alpha)\\
\Rightarrow \ln\left(e^{\Delta G_{new}^{o}}\right) &= \ln\left(e^{\Delta G_{ori}^{o}}(1-\alpha)\right)\\
\Rightarrow \Delta G_{new}^{o} &= \Delta G_{ori}^{o}+\ln\left(1-\alpha\right)\\
\Rightarrow \Delta\left(\Delta G^{o}\right) &= \ln\left(1-\alpha\right)\\
\end{aligned}
\label{eq:rateconversion}
\end{equation}
Equation~\ref{eq:rateconversion} gives us the change in energy corresponding to a relative change \textalpha. Practically, when computing derivatives such as the ones in equation for the Fisher Information Matrix, we thus use a relative rate of \textalpha for kinetic rates, and of $\ln\left(1-\alpha\right)$ for energies. Notice that as \textalpha goes to $0$ the relative change of energy also becomes zero. However, we can not pick an \textalpha that is too small due to numerical imprecision. Figure~\ref{fig:MT_Rateconversion} shows the relationship between the change in energy and the ratio of change in rates. For example, a 5$\%$ change in rate constant is around $0.05$ $k_{B}T$ change in energy.

\vspace{-10pt}
\section*{Appendix S3: Analytic calculation of eigenvalues for a one dimensional diffusion system at the first time step}
\vspace{-5pt}
\noindent For the one dimensional random walk where the particles diffuse uniformly, the probability density after one time step is proportional to the number of particles at each possible lattice site. Therefore, $P(N_{\mu}) = \frac{N_{\mu}}{N_{total}}=\kappa$ where $N_{\mu}$ is the number of particles that are in each possible lattice sites and total number of particles $N_{total} = \sum_{\mu} N_{\mu}$.

Using equation~\ref{eq:FIMexp} from Appendix S1 and taking into account that after the first time step only the diagonal elements of the Fisher Information Matrix are nonzero. Therefore, equation~\ref{eq:FIMexp} simplifies to:
\begin{equation}
g_{\mu,\nu} =  \frac{\partial y}{\partial \theta_{\mu}} \frac{\partial y}{\partial \theta_{\nu}}\theta_{\mu}\theta_{\nu}\delta_{\mu\nu}
\end{equation}
where $\delta_{\mu\nu}$ is the Kronecker delta.

\begin{figure*}[!htb]
  \begin{minipage}[c]{0.67\textwidth}
    \centering
    \includegraphics[width=\textwidth]{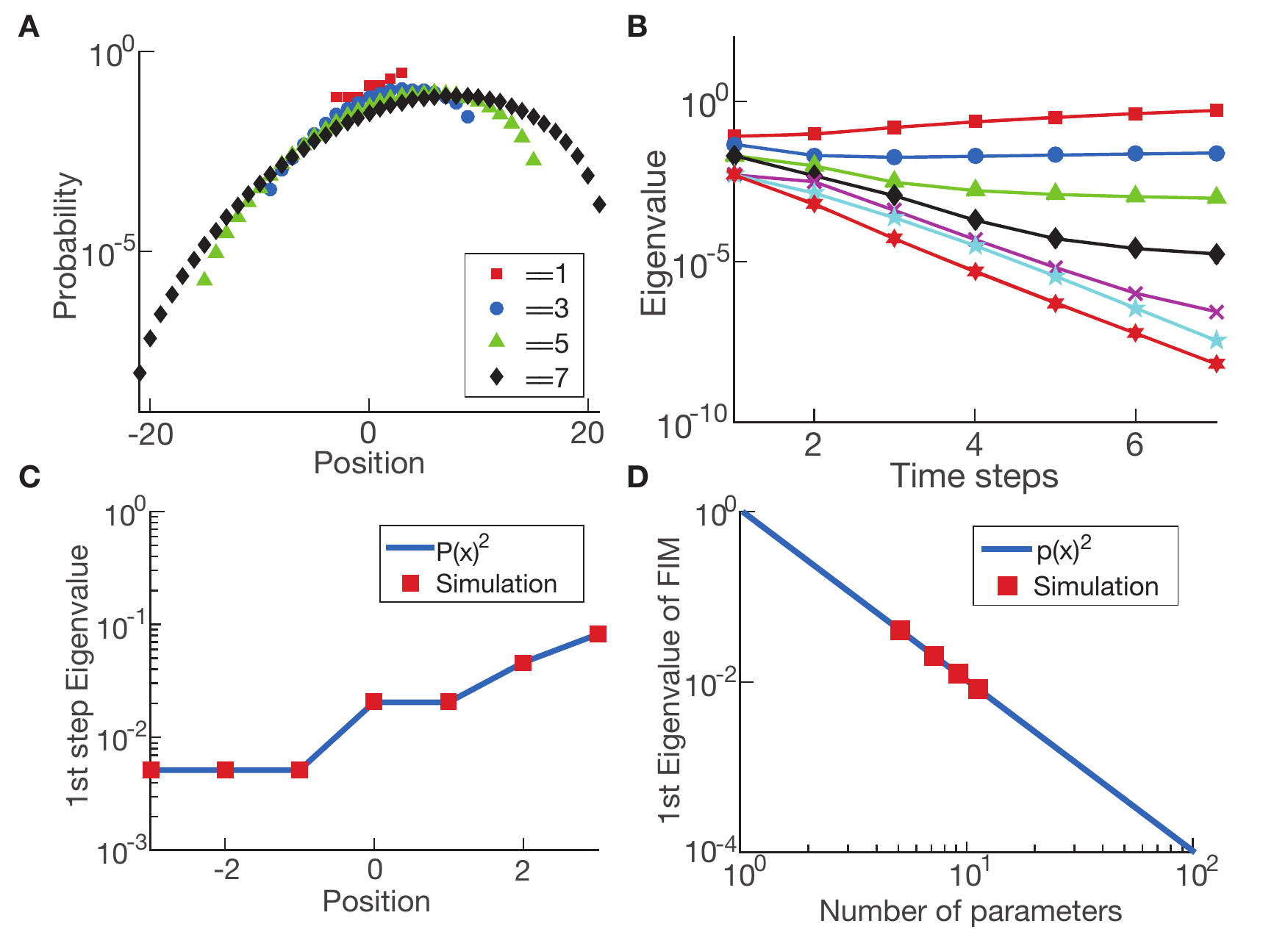}
  \end{minipage}
  \begin{minipage}[c]{0.3\textwidth}
  \vspace{-20pt}
    \caption{\fontencoding{T1}\fontfamily{phv}\selectfont{\bf Figure S\ref*{fig:RW_combine}.} (A) The particle density for a non uniform one dimensional random walk over time (with probability higher to the right side of the space). (B) The eigenvalues for the non uniform one dimensional random walk. The eigenvalues at the first step of the simulation is not unity compare to the uniform random walk. (C) At the first step of the simulation for the one dimensional random walk, the eigenvalue is equal to the squared rate given for the non uniform simulation. This result is universal for any rates given. (D) The eigenvalues at the first step of a uniform simulation is also equal to the squared rate of the simulation \textit{i.e.} squared of one over the number of parameters.}
\label{fig:RW_combine}
  \end{minipage}
\end{figure*}

\noindent Therefore, the corresponding eigenvalues are:
\begin{equation}
	\begin{aligned}
	    \lambda_{\mu} &= g_{\mu,\mu} =  \left[\frac{y_{\theta+\Delta\theta}-y_{\theta-\Delta\theta}}{2\Delta\theta}\right]^2\theta^2 \\
		&= \left[\frac{\frac{N_{\mu}+\Delta N_{\mu}}{N_{tot}}-\frac{N_{\mu}-\Delta N_{\mu}}{N_{tot}}}{\frac{2\Delta N_{\mu}}{N_{tot}}}\right]^2\\
		&= \left[(N_{\mu}+\Delta N_{\mu})-(N_{\mu}-\Delta N_{\mu})\right]^2 \frac{N_{\mu}^2}{4\Delta N_{\mu}^2N_{tot}^2}\\
		&= (2\Delta N_{\mu})^2\left( \frac{N_{\mu}^2}{4\Delta N_{\mu}^2N_{tot}^2}\right)\\
		&= \frac{N_{\mu}^2}{N_{tot}^2}\\
		&= P(N_{\mu})^2\\
	\end{aligned}
\end{equation}.
\noindent This shows that the eigenvalue of a one dimensional random walk at the first time step is equal to the probability at a given lattice sites squared. The result for both the drift right diffusion and uniform diffusion with different number of sites are shown in figure~S\ref{fig:RW_combine}C and D.

\vspace{-10pt}
\section*{Appendix S4: Analytic calculation of the dominant eigenvalue for the protein production and degradation system}
\vspace{-5pt}
\noindent The stationary distribution of a simple production and degradation system is given by the Poisson distribution~\cite{Systembiology_2016}:
\begin{equation}
P(\rho,\delta,n) = \frac{1}{n!}e^{-\frac{\rho}{\delta}}\left(\frac{\rho}{\delta}\right)^n
\label{eq:poisson}
\end{equation}
therefore, the Fisher Information Matrix for this system from equation~\ref{eq:FIMexp} becomes:
\begin{equation}
\begin{gathered}
\begin{bmatrix}
\sum_{n} \left(\frac{\partial P_{n}}{\partial\rho}\rho\right)^2 & \sum_{n} \frac{\partial P_{n}}{\partial\rho}\rho\frac{\partial P_{n}}{\partial\delta}\delta\\
 \sum_{n} \frac{\partial P_{n}}{\partial\rho}\rho\frac{\partial P_{n}}{\partial\delta}\delta & \sum_{n} \left(\frac{\partial P_{n}}{\partial\delta}\delta\right)^2
\end{bmatrix}
\\=
\begin{bmatrix}
\sum_{n} \alpha_{n}^2 & \sum_{n} \alpha_{n}\beta_{n} \\
\sum_{n} \alpha_{n}\beta_{n} & \sum_{n} \beta_{n}^2 \\
\end{bmatrix}
\end{gathered}
\end{equation}
where $\alpha_{n} = \frac{\partial P_{n}}{\partial\rho}\rho$ and $\beta_{n}=\frac{\partial P_{n}}{\partial\delta}\delta$.

\noindent To find the eigenvalues $\lambda$ of this matrix, we subtract the identity matrix with diagonal value $\lambda$ and set the determinant equals to zero:
\begin{equation}
\det
\begin{bmatrix}
\sum_{n} \alpha_{n}^2-\lambda & \sum_{n} \alpha_{n}\beta_{n} \\
\sum_{n} \alpha_{n}\beta_{n} & \sum_{n} \beta_{n}^2-\lambda \\
\end{bmatrix}
= 0
\end{equation}

\noindent We can calculate $\alpha_{n}$ and $\beta_{n}$ analytically and show that the magnitudes of the two terms are equal to each other:
\begin{equation}
 \alpha_{n} = |\beta_{n}| = P(n)\frac{(\rho-n\delta)}{\delta}
\end{equation}
Thus the determinant matrix becomes the following form, where $A=-\alpha_{n}\beta_{n}$:
\begin{equation}
\det
\begin{bmatrix}
A-\lambda & -A \\
-A & A-\lambda \\
\end{bmatrix}
= 0
\end{equation}
The eigenvalues for this matrix are easy to compute with only one nonzero eigenvalue:
\begin{equation}
\lambda_{1} = 2A = 2\sum_{n} P(n)^2\frac{(\rho-n\delta)^2}{\delta^2}
\end{equation}

To calculate analytically the nonzero eigenvalue: $\lambda_{1}$, we approximate the Poisson distribution with a Gaussian distribution and change the integral into a summation with $\gamma=\frac{\rho}{\delta}$:
\begin{equation}
\begin{aligned}
\lambda_{1}  &= 2\sum_{n} P^2\frac{(\rho-n\delta)^2}{\delta^2}\\
			&\simeq 2\int_{0}^{\infty} \left(\frac{1}{\sqrt{2\pi\gamma}}e^{\frac{(x-\gamma)^2}{2\gamma}}\right)^2\left(\frac{(\rho-x\delta)^2}{\delta^2}\right) dx \\
			&= \frac{1}{\pi}\left[\frac{1}{4}\sqrt{\pi\gamma}erf\left(\frac{x}{\sqrt{\gamma}}-\sqrt{\gamma}\right)+\frac{1}{2\delta}e^{-\frac{(x-\gamma)^2}{\gamma}}(\rho-\delta x)\right]_{0}^{\infty}\\
            &= \frac{1}{2}\sqrt{\frac{\gamma}{\pi}}-\frac{\gamma}{2\pi}e^{-\gamma}
\end{aligned}
\end{equation}
\begin{figure*}[!tbh]
\centering
\includegraphics[width=0.67\textwidth]{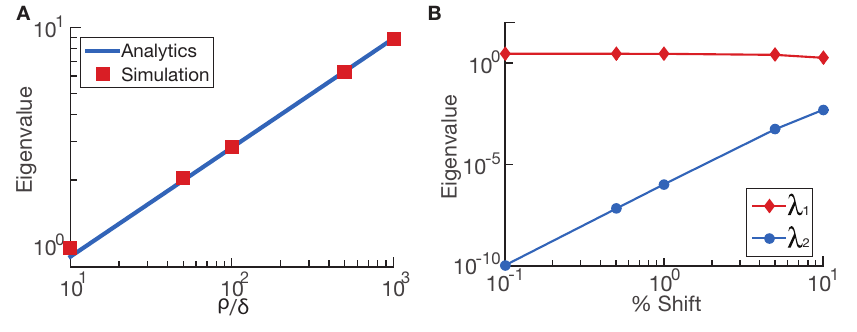}
\caption{\fontencoding{T1}\fontfamily{phv}\selectfont{\bf Figure S\ref*{fig:Protein_combine}.} (A) The dominating dominating eigenvalue for the protein production degradation system is shown to be a ratio of the production rate over the degradation rate times some constant. The simulations of different ratios matches the analytic solution. (B) The second eigenvalues for the protein production and degradation rate is non zero during simulation due to the limitation of the physical system itself. At steady state the average number of protein is one hundred protein which means that the smallest shift for probability to calculate the finite derivatives is one protein which is one percent which has a nonzero eigenvalue.}
\label{fig:Protein_combine}
\end{figure*}
If the value of $\gamma\gg0$, the second term becomes negligible due to the exponential. Thus, the final expression for the eigenvalue is:
\begin{equation}
\lambda_{1} \simeq \frac{1}{2}\sqrt{\frac{\gamma}{\pi}} = \frac{1}{2}\sqrt{\frac{\rho}{\delta\pi}}
\end{equation}

\noindent It is important to note that even though the second eigenvalue is zero from the analytic calculation, due to the limitation of the numerical precision, it is impossible to reach the zero value given the physical definition of the system. For example if production rate $\rho=1$ and the degradation rate $\delta=0.01$, we know that the steady state solution will have a peak at number of protein of $N=100$. However, this means that when calculating the finite derivatives, using any shift smaller than $1\%$ will result in the shift being less than one protein which is nonphysical. We show that even using complete analytic values of Poisson distribution to calculate the finite derivatives, we will not be able to reach zero when using the finite derivatives. The figure is shown in Appendix figure~S\ref{fig:Protein_combine}B.

\vspace{-10pt}
\section*{Appendix S5: Microtubule simulation}
\vspace{-5pt}
We use VanBuren \textit{et al.} 2002 as our inspiration for our simulation of microtubule dynamic  instability~\cite{VanBuren_Odde_2002}.
\vspace{-10pt}
\subsection*{Parameters}
\vspace{-5pt}
The microtubule has 13 protofilaments with a three monomer offset at the seam. The base parameters in the model are: (1) $k_{+}$, the association rate constant for tubulin subunts to associate with the end of a protofilament; (2) $\Delta G^{o}_{long}$, the longitudinal bond energy between dimers; (3) $\Delta G^{o}_{lat}$, the lateral bond energy between tubulin subunits in a B-lattice configuration ($\alpha-\alpha$ and $\beta-\beta$); (4) $k_{H}$, the hydrolysis rate constant for the conversion of GTP-tubulin to GDP-tubulin; (5) $\Delta\Delta G_{lat}^{o}$, the change in free energy associated with GTP hydrolysis, which is assigned to each lateral bond; and (6) $[Tubulin]$, the concentration of tubulin. The additional parameters added to the model are: (7) $\Delta G_{seam}^{o}$, the lateral bond energy between tubulin subunits in an A-lattice configuration, namely at the seam ($\alpha-\beta$ and $\beta-\alpha$); and (8) $k_{s}$, the dimensionless factor that reduces $k_{+}$ in the presence of lateral neighbors. See Fig.~\ref{fig:MT_system}A and B for schematics.
\vspace{-10pt}
\subsection*{Coupled Random Hydrolysis}
\vspace{-5pt}
An \textalpha-tubulin contributes catalytic residues to the GTP pocket of the \textbeta-tubulin that sits below it. Therefore, a tubulin subunit cannot hydrolyze its GTP unless another tubulin subunit is above it; in other words, only non-terminal subunits can hydrolyze GTP. The GTP hydrolysis reaction for all non-terminal subunits occurs at random time intervals (see Gillespie algorithm below). This implementation of GTP hydrolysis is known as coupled random hydrolysis~\cite{Bowne-Anderson_Howard_2013}. We do note, however, that earlier models of dynamic instability used other implementations of GTP hydrolysis, e.g., where the hydrolysis reaction was obligate after association of a new terminal subunit, known as vectorial hydrolysis. But coupled random hydrolysis is the standard among contemporary models.

Our model assumes the transition from GTP-tubulin to GDP-tubulin occurs in a single step, with no intermediates in the hydrolysis pathway. A single step is surely a simplification, as {GTPases} often have important GDP-Pi intermediate states. Recently, Manka et al. solved a cryo-EM structure of the putative GDP-Pi state~\cite{Manka_Moores_2018}, and very recent models have incorporated a GDP-Pi state explicitly~\cite{Kim_Rice_2018}. Testing the relevance of a GDP-Pi state will be the subject of future studies.
\begin{figure*}[!thb]
\centering
\includegraphics[width=0.67\textwidth]{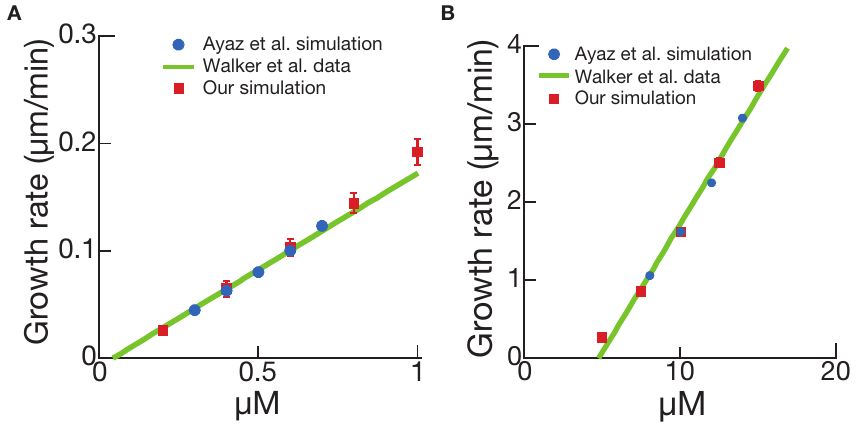}
\caption{\fontencoding{T1}\fontfamily{phv}\selectfont{\bf Figure S\ref*{fig:MT_Benchmark_combine}.} Microtubule simulation bench mark against computer simulation of Ayaz \textit{et al.} and experimental data of Walker \textit{et al.} (A) GMPCPP tubulin (B) GTP tubulin.}
\label{fig:MT_Benchmark_combine}
\end{figure*}
\vspace{-10pt}
\subsection*{Lateral Weakening}
\vspace{-5pt}
The effect of GTP hydrolysis in the VanBuren model is a weakening of lateral bonds. This choice is based on the observation that protofilaments peel outward after a catastrophe ~\cite{Mandelkow_Milligan_1991}, which indicates that the lateral bonds rupture first. More recent observations suggest that longitudinal bonds may also be affected by GTP hydrolysis; more specifically, the N-domain of \textalpha-tubulin appears to compact down into the \textbeta-tubulin below it.
\vspace{-10pt}
\subsection*{Gillespie algorithm}
\vspace{-5pt}
We use the direct method of the Gillespie algorithm to simulate all the possible events for the microtubule at a given time~\cite{Gillespie_1977}. The possible events for the microtubule simulation are: association of dimers, dissociation of dimers, and hydrolysis of GTP-tubulin to GDP-tubulin. The association and dissociation events only happen at the ends of protofilaments. The association rate is equal on the top of each protofilament and is defined as the association constant $k_{+}$ multiplies by the concentration of tubulin:
\begin{equation}
k_{on,PF} = k_{+}[Tubulin]
\end{equation}
The dissociation rate depends on $\Delta G_{total}^{o}$: the total bond energies the tubulin dimer have with its neighboring dimers:
\begin{equation}
k_{off} = \frac{k_{+}}{e^{-\frac{\Delta G_{total}^{o}}{k_{B}T}}}
\end{equation}
The total number of hydrolysis events depend on the number of GTP tubulin that are present in the lattice and a nonterminal GTP can be hydrolyzed into a GDP at a fixed first order rate $k_{H}$ which leads to a total rate of hydrolysis at any given time as:
\begin{equation}
k_{hyd} = k_{H}\times N(GTP)
\end{equation}
The effect of hydrolysis is a decrease in the bond strength laterally by $\Delta\Delta G^{o}$. This is commonly known as the coupled random hydrolysis model.

 We sum up all the possible rates $r_{i}$ for all the possible events \textalpha and generate two random number $R_{1}$, $R_{2}$ between zero and one. We choose the events that satisfy the following condition:
\begin{equation}
\frac{\sum^{i-1}_{0} r_{i}}{\alpha} \leq R_{1} < \frac{\sum^{i}_{0} r_{i}}{\alpha}
\end{equation}
and we increment the simulation time $t$ by $\tau$, where $\tau$ is defined as
\begin{equation}
\tau=\frac{1}{\alpha}\log\left(\frac{1}{R_{2}}\right)
\end{equation}
\vspace{-10pt}
\subsection*{Benchmarking}
\vspace{-5pt}
Because we use the direct method of the Gillespie algorithm, we benchmarked our Gillespie simulation against published results of Ayaz \textit{et al.}~\cite{Ayaz_Luke_2014} and VanBuren \textit{et al.}~\cite{VanBuren_Odde_2002} to ensure that our simulation was properly implemented. We used their precise parameter values and simulated microtubule growth in the absence of GTP hydrolysis (which is handled differently in the two models). Our simulation recovered their results exactly (see Fig.~\ref{fig:MT_Benchmark_combine}A and~\ref{fig:MT_Benchmark_combine}B). Therefore our simulation is well-executed.
\vspace{-10pt}
\subsection*{Reproduction of Experimental Data}
\vspace{-5pt}
Simple models like those described here are best at reproducing microtubule growth rates and post-catastrophe shrinkage rates across a range of tubulin concentrations. The models can reproduce the mean lifetime at a single tubulin concentration, but then the trouble begins. These models cannot reproduce the mean lifetime across a range of tubulin concentrations. The models are much too sensitive to $[Tubulin$--at high $[Tubulin$, catastrophes become exceedingly rare, in contrast with experimental observations. Similarly, the distribution of microtubule lifetimes is best described by a Gamma distribution because catastrophes are a result of microtubule ``aging''. Simple models do not reproduce this Gamma distribution but rather given an exponential distribution of lifetimes. More complex models do better. At present, no published model is able to reproduce data on templated nucleation.

\vspace{-10pt}
\section*{Appendix S6: Microtubule simulation analysis}
\vspace{-5pt}
We use an in house Matlab code to analyze the microtubule trajectories over time. The algorithm smooth the trajectories and identify local maxima and minima. The maxima correspond to potential catastrophe positions and minima correspond to potential rescue positions. lifetime and post-catastrophe shrinkage rate is then determined. Note: to avoid fluctuation of the stochastic simulation, a threshold of $250$ $nm$ is used (since this is the point spread function of our microscope) such that any lifetime that is counted towards the probability distribution starts from a minima to a maxima while passing $250$ $nm$ and vice versa for the post-catastrophe shrinkage rate but instead it goes from a maxima to a minima.
\begin{figure*}[!htb]
  \begin{minipage}[c]{0.67\textwidth}
    \centering
    \includegraphics[width=\textwidth]{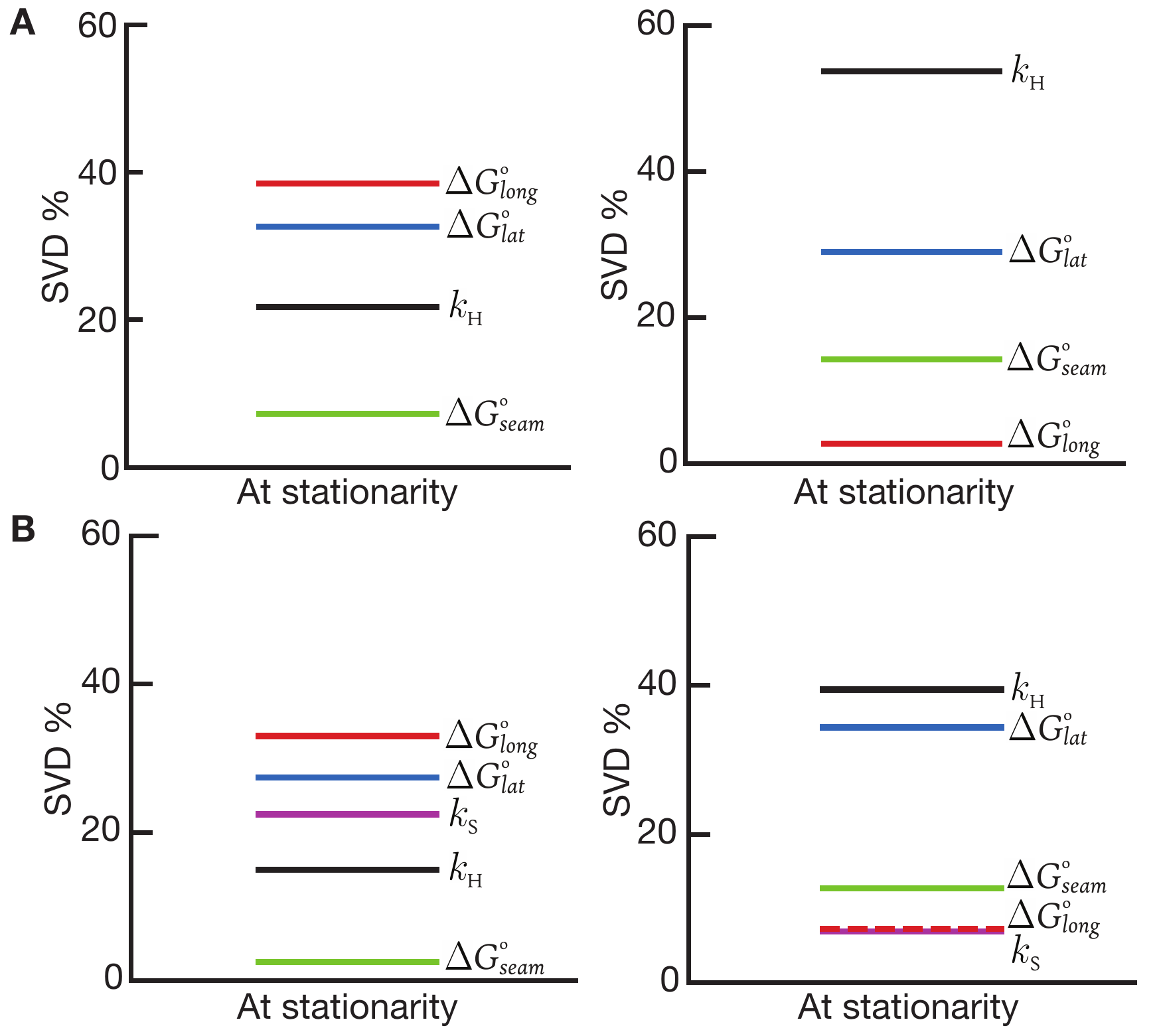}
  \end{minipage}
  \begin{minipage}[c]{0.3\textwidth}
  \vspace{-120pt}
    \caption{\fontencoding{T1}\fontfamily{phv}\selectfont{\bf Figure S\ref*{fig:SVD_combine}.} The first column is the leading vector for the highest singular value for the microtubule system. The longitudinal bond and lateral bond are the controlling parameters. The second column is the leading vector for the second highest singular value for the microtubule system. The lateral bond and the hydrolysis rate is the controlling parameters: (A) Microtubule simulation without neighboring penalty. (B) Microtubule simulation with neighboring penalty Note: The SVD $\%$ plotted are the absolute value of parameter component.}
\label{fig:SVD_combine}
  \end{minipage}
\end{figure*}

\newpage
\printbibliography[title={APPENDIX REFERENCES}]
\end{refsection}

\end{multicols}

\end{document}